\documentclass[prb,twoside,twocolumn,10pt,floatfix,showpacs,citeautoscript]{revtex4-2}

\usepackage{amsmath}
\usepackage{amssymb}
\usepackage{color}
\usepackage{glossaries}
\usepackage{graphicx}
\usepackage{times}
\usepackage{units}
\usepackage[version=4]{mhchem}
\usepackage{siunitx}

\usepackage[
    unicode,
    colorlinks=true,
    urlcolor=blue,
    linkcolor=blue,
    citecolor=blue
]{hyperref}

\graphicspath{{./figures/}}

\newacronym{aic}{AIC}{Akaike information criterion}
\newacronym{bic}{BIC}{Bayesian Information Criterion}
\newacronym{bz}{BZ}{Brillouin zone}
\newacronym{ce}{CE}{cluster expansion}
\newacronym{dft}{DFT}{density-functional theory}
\newacronym{eci}{ECI}{effective cluster interaction}
\newacronym{fcc}{FCC}{face-centered cubic}
\newacronym{fu}{f.u.}{formula unit}
\newacronym{lro}{LRO}{long-range order}
\newacronym{lrtddft}{LR-TDDFT}{linear-response time-dependent density-functional theory}
\newacronym{mc}{MC}{Monte Carlo}
\newacronym{mcmc}{MCMC}{Markov chain Monte Carlo}
\newacronym{sos}{SOS}{special ordered structure}
\newacronym{sgc}{SGC}{semi-grand canonical}
\newacronym{sro}{SRO}{short-range order}
\newacronym{vcsgc}{VCSGC}{variance-constrained semi-grand canonical}

\begin{document}

\title{A tale of two phase diagrams: Interplay of ordering and hydrogen uptake in Pd--Au--H}

\author{J. Magnus Rahm}
\author{Joakim Löfgren}
\author{Erik Fransson}
\author{Paul Erhart}
\email{erhart@chalmers.se}
\affiliation{
  Chalmers University of Technology,
  Department of Physics,
  S-412 96 Gothenburg, Sweden
}
\begin{abstract}
Due to their ability to reversibly absorb/desorb hydrogen without hysteresis, Pd--Au nanoalloys have been proposed as materials for hydrogen sensing.
For sensing, it is important that absorption/desorption isotherms are reproducible and stable over time.
A few studies have pointed to the influence of short and long range chemical order on these isotherms, but many aspects of the impact of chemical order have remained unexplored.
Here, we use alloy cluster expansions to describe the thermodynamics of hydrogen in Pd--Au in a wide concentration range.
We investigate how different chemical orderings, corresponding to annealing at different temperatures as well as different external pressures of hydrogen, impact the behavior of the material with focus on its hydrogen absorption/desorption isotherms.
In particular, we find that a long-range ordered L1$_2$ phase is expected to form if the \ce{H2} pressure is sufficiently high.
Furthermore, we construct the phase diagram at temperatures from \unit[250]{K} to \unit[500]{K}, showing that if full equilibrium is reached in the presence of hydrogen, phase separation can often be expected to occur, in stark contrast to the phase diagram in para-equilibrium.
Our results explain the experimental observation that absorption/desorption isotherms in Pd--Au are often stable over time, but also reveal pitfalls for when this may not be the case.
\end{abstract}

\maketitle

\section{Introduction}
The prospect of hydrogen as a replacement for fossil fuels continues to generate interest in metallic hydrides.
Some metals are of particular interest in this context as they can be reversibly loaded and unloaded with hydrogen by tuning the partial pressure of \ce{H2} gas in the environment; Pd is one such example.
When loaded with hydrogen, the Pd hydride has a hydrogen density that is orders of magnitudes larger than \ce{H2} gas under the same conditions, and it is therefore a candidate for storing hydrogen \cite{SchWhiKan18}.
Another potential application of Pd hydrides is sensing \cite{WadSyrLan14, DarNugLan20}.
Pd nanoparticles that are exposed to hydrogen quickly form a hydride \cite{NarHayBal17}, and when doing so, their optical properties change.
This change can be easily detected, and Pd nanoparticles can thus be used as reversible hydrogen sensors \cite{NugDarCus19, DarStoOst20}.
The use of Pd for hydrogen sensing applications has, however, a major disadvantage; Pd hydride formation at room temperature is associated with a first-order phase transition from a hydrogen-poor $\alpha$ phase to a hydrogen-rich $\beta$ phase, which causes the sensor response to be a highly non-linear function of \ce{H2} pressure.
Since the phase transition is also associated with hysteresis \cite{FlaOat91, GriStrGie16, SyrWadNug15}, the response moreover differs between the loading and unloading half-cycles.
These aspects are generally unfavorable for sensing applications.

Fortunately, these drawbacks can be alleviated by alloying Pd with Au.
With around 20\,mol-\% of Au in Pd, the system can be loaded continuously with hydrogen \cite{MaeFla65, LuoWanFla10}, eventually making the sensor readout an almost linear function of \ce{H2} pressure \cite{WadNugLid15, NugDarZhd18, WesRooLec13, BanNugSch19}.
The introduction of another chemical species thus fundamentally changes the thermodynamics of the material and improves its properties.
From a scientific as well as practical standpoint, however, an alloy is significantly more complex than a pure metal, and multiple important questions arise: what is the optimal composition, how are the chemical elements ordered under different circumstances, and how does this ordering influence the properties of the material?
The influence of overall composition can usually be relatively easily optimized by experimental screening.
The chemical ordering, on the other hand, is difficult to both control and measure experimentally, and can be expected to be influenced by conditions during and after fabrication.
Moreover, although Pd--Au nanoalloy sensors tend to be stable over time \cite{PakJeoYeo19}, changes may occur over long time-scales due to slow kinetics, which makes the effects tedious to experimentally assess, while they may still be detrimental to the material.
Although the study of hydrogen in Pd and its alloys has a long and rich history \cite{Ber11, Sch32, FlaOat91, SakCheUra95, Fuk06, PunKir06, HuaOpaWan07, JouThi11}, many fundamental aspects of the Pd--Au--H system are still unknown, and to fully exploit this material it is of paramount importance that its thermodynamics are understood, as it determines the driving forces that underpin the evolution of the atomic scale structure over time.

A few previous studies have successfully used combinations of \gls{dft} calculations and statistical methods to predict properties of hydrogen in Pd--Au.
For example, Mamatkulov and Zhdanov \cite{MamZhd20} recently developed a model based on \gls{dft} calculations that successfully predicted absorption isotherms, highlighting that once octahedral sites without surrounding Au atoms become few, the phase transition and its associated hysteresis disappears.
These calculations assumed a Pd--Au lattice with random configuration, which remained unaffected by insertion of hydrogen.
This metastable equilibrium, with equilibrium only on the hydrogen sublattice, is commonly referred to as para-equilibrium \cite{Hul47, HilAgr04, FlaOat19}.
If the metal atoms are allowed to rearrange in response to hydrogen such that full equilibrium is reached, a significantly more complex phase diagram can be expected \cite{NanDenBot06}.
This was indicated experimentally by Lee \textit{et al.} \cite{LeeNohFla07}, who found that \gls{lro} was formed in Pd--Au with 19\% Au when subjected to a high pressure of \ce{H2} at high temperatures.
Importantly, alloys with this \gls{lro} had strikingly different pressure--composition isotherms than those that had not been treated in a high-pressure \ce{H2} gas and lacked \gls{lro}.
A similar conclusion was reached by Chandrasekhar and Sholl \textit{et al.} \cite{ChaSho14}, who computed hydrogen solubility as a function of chemical order at Au contents 4\% and 15\%.
These results show that the different degrees of equilibration matter for the properties of the Pd--Au--H system,
and since many applications depend on reproducible pressure--composition isotherms, chemical order should not be ignored.
In this light, it is obvious that a full accounting of the impact of chemical order on the thermodynamics of Pd--Au hydride is desirable.

Here, we make a detailed atomic scale study of the thermodynamics of hydrogen in Pd--Au.
To this end, we use alloy \glspl{ce} fitted to \gls{dft} to describe the energetics of the system ranging from 0\% to 50\% Au in Pd and in the full range from no hydrogen to 100\% hydrogen (i.e., 1 H atom per 1 Au or Pd atom).
This approach, which has been successfully applied to similar materials in the past \cite{SemSho08, ChaSho14, BouCenCri19, GunHarVen18, OuyChaKim19}, allows for very fast evaluation of the energy of a system with hundreds of atoms with an accuracy approaching that of \gls{dft}.
Using \gls{mc} simulations, we can elucidate the thermodynamics and, in particular, study how chemical order evolves under different circumstances.

\section{Methodology}

\subsection{Structure generation}

In Pd--Au hydride, hydrogen primarily occupies octahedral interstitial sites in the \gls{fcc} lattice formed by Pd--Au (although in some configurations, tetrahedral site may be preferred \cite{MamZhd20, NanLegEij08}).
The full system can thus be viewed as a rock salt structure, i.e., two intertwined \gls{fcc} sublattices, one occupied by Pd and Au, and the other populated with H and vacancies.
In the following, we will state concentrations \emph{per sublattice}, such that, for example, the H concentration is 100\% if all sites on the H/vacancy sublattice are occupied by H, whereas energies are stated per \gls{fu} with a \gls{fu} being two sites, one per sublattice.

Structures were generated by exhaustive enumeration \cite{AngMunRah19, HarFor08} of supercells of the primitive structure with up to 12 atoms in the cell (6 per sublattice).
The limited supercell size precludes structures with Au concentration between 0 and 16.67\% but as this is a composition interval of particular interest, we generated an additional 200 structures in this interval having up to 32 atoms in the cell, heuristically chosen so as to properly span the space of cluster vectors.
Finally, in order to properly describe the important limit of pure Pd hydride, we generated 367 structures without Au, having up to 24 atoms in the cell.

\subsection{DFT calculations}
For generation of training data, we used the projector augmented wave formalism as implemented in the Vienna ab initio simulation package (version 5.4.1, PAW 2015) \cite{KreFur96b, KreJou99} with the vdW-DF-cx exchange-correlation functional \cite{DioRydSch04, BerHyl14}.
Wave functions were expanded in a plane wave basis set with a cutoff of \SI{400}{\eV}, the \gls{bz} was sampled with a $\Gamma$-centered grid with a $\mathbf{k}$ point spacing of \SI{0.2}{\per\angstrom}, and occupations were set using Gaussian smearing with a width of \SI{0.1}{\eV}.
Atomic positions and cell shape were relaxed until residual forces were below \SI{25}{\milli\eV/\angstrom} and stresses below \SI{1}{\kilo\bar}.
Once converged, we ran an additional single-point calculation with a $\mathbf{k}$ point spacing of \unit[0.1]{\AA$^{-1}$} and the \gls{bz} was integrated using the tetrahedron method with Bl\"{o}chl corrections.

\subsection{Alloy cluster expansions}
In an alloy \gls{ce}, the energy of a configuration of atoms $\mathbf{\sigma}$ is decomposed into contributions from single sites (``singlets''), pairs of sites, triplets of sites, etc, commonly referred to as clusters.
Denoting such clusters $\mathbf{\alpha}$, the energy can be expressed as
\begin{equation} \label{eq:ce}
    E(\mathbf{\sigma}) = J_0 + \sum_\mathbf{\alpha} J_\mathbf{\alpha} m_\mathbf{\alpha} \left \langle \Pi_\mathbf{\alpha}(\mathbf{\sigma}) \right \rangle_\mathbf{\alpha},
\end{equation}
where the \gls{eci} $J_\mathbf{\alpha}$ measures the energy associated with the cluster $\mathbf{\alpha}$ (with $J_0$ being a constant, the ``zerolet''),
$m_\mathbf{\alpha}$ is the number of symmetrically equivalent clusters $\mathbf{\alpha}$,
and  $\left \langle \Pi_\alpha(\mathbf{\sigma}) \right \rangle_\mathbf{\alpha}$ quantifies the chemical ordering of the clusters $\mathbf{\alpha}$.
Specifically, $\left \langle \Pi_\alpha(\mathbf{\sigma}) \right \rangle_\mathbf{\alpha}$ is an average over all symmetrically equivalent clusters $\mathbf{\alpha}$, where each term is a product of so-called point functions,
\begin{equation}
    \Pi_\alpha(\mathbf{\sigma}) = \prod_{i \in \alpha} \Theta_i,
\end{equation}
meaning that the product runs over all lattice sites $i$ in the cluster $\mathbf{\alpha}$.
Here, we explicitly treat unoccupied interstitial sites as vacancies, which makes it possible to view the interstitial sublattice as an alloy.
In our case, the point functions then take the value $\Theta_i=-1$ if site $i$ is occupied by Au or H, or $\Theta_i=1$ if site $i$ is occupied by Pd or a vacancy (but the sublattices are distinct, meaning that, e.g., hydrogen cannot occupy a site on the Pd--Au sublattice).

Although it can be shown that this basis set forms a complete orthonormal basis and thus can be used to exactly express any function of the configuration $\mathbf{\sigma}$ \cite{SanDucGra84}, in practice it has to be truncated.
Physical intuition tells us that clusters with few sites and short interatomic distances should be more important, and by analyzing the \gls{aic} and \gls{bic} \cite{ZhaLiuBi20} (\autoref{sfig:rmse-aic-bic}), we found that it was sufficient to include pairs no more than 1.75$a_0$ apart (where $a_0$ is the lattice parameter of the conventional (cubic) cell) and triplets that had all interatomic distances shorter than 1.25$a_0$.
With these cutoffs, we end up with 18 and 42 symmetrically distinct pairs and triplets, respectively.

The \glspl{eci} $J_\alpha$ were deduced by fitting the linear equation \eqref{eq:ce} to the energies calculated with \gls{dft} using a Bayesian approach \cite{MueCed09}.
We assumed that the errors were normally distributed with variance $\sigma_1^2$, and used an inverse gamma function as prior for $\sigma_1$.
Furthermore, we used Gaussian priors for the \glspl{eci} (having mean $\mu = 0$ and variance $\sigma_2^2$), and used an inverse gamma function as prior for $\sigma_2$.
(We note that the fitting process was largely insensitive to the choice of priors due to the large training data set.)
We then sampled the posterior distribution using \gls{mcmc} simulations as implemented in the \textsc{emcee} software \cite{ForHog13} with 300 walkers and 417,000 MC steps (100 times the auto-correlation length).
From the resulting posterior distributions, we constructed our final model as the average over the posterior distribution for each \gls{eci} (we used the average rather than the peak of the posterior since the posteriors were almost perfectly symmetric and the two measures therefore coincided to within less than \unit[0.5]{meV/\gls{fu}}, and the average is insensitive to noise and choice of bins).
All of the results presented below were obtained with this \gls{ce} unless otherwise stated.
In addition to this \gls{ce}, we randomly selected 10 independent \glspl{ce} from the \gls{mcmc} trajectory, which were used to investigate the impact of variations in the \glspl{eci} on thermodynamic properties of the model.
For details on priors and posterior distributions, see \autoref{sfig:hyper-priors-and-posteriors}--\ref{sfig:ecis-corner}.

It is usually practical to fit \glspl{ce} to mixing energies, defined such that the pure phases have zero mixing energy.
In our system, we have four pure phases: Au and Pd with and without hydrogen.
Only linear transformations are permissible from a thermodynamic perspective since only first-order terms cancel.
This implies that only three (out of four) boundary points can be set strictly to zero.
Here, we make the (intuitively logical) choice to set Pd, PdH and Au to zero, leaving AuH using the following definition of the mixing energy:
\begin{align}
    \begin{split}
    E_\text{mix, DFT}(\mathbf{\sigma}) =& E_\text{DFT}(\mathbf{\sigma}) / N
      -\left[c_\text{Pd} - c_\text{H}\right] E_\text{DFT}(\text{Pd})\\
     &- c_\text{H} E_\text{DFT}(\text{PdH})
     - c_\text{Au} E_\text{DFT}(\text{Au})
    \end{split},
\end{align}
where $N$ is the number of \glspl{fu} in the structure (i.e., the total number of Au and Pd atoms).

Generation of structures with exhaustive enumeration always yields disproportionately many structures close to 50\% concentrations, as it corresponds to the largest number of symmetrically distinct structures.
To counter this bias, we constrained the \gls{ce} to always reproduce the mixing energy of Pd, PdH, and Au to their exact mixing energy values (\unit[0]{meV/\gls{fu}} per definition).

\subsection{Monte Carlo simulations}

\Gls{mc} simulations were carried out according to a standard Metropolis scheme, in which the chemical identity of one or more atoms is changed in each step, and the change is accepted with probability $P = \min \left\lbrace 1, \exp{\left(-\Delta \psi / k_\text{B} T\right)} \right \rbrace$.
The change in the potential $\Delta \psi$ can be modified to account for different physical situations.
Here, we used three different approaches.

The first approach is to sample the canonical ensemble by letting $\psi$ be the change in potential energy, $\psi = E$, and always swap the identities of two unlike atoms on the same sublattice.
This has the advantage of preserving the overall concentration of the chemical species and is thus reminiscent of typical experimental conditions for the Pd--Au lattice.
In the case of hydrogen, however, one experimentally controls the chemical potential rather than the concentration, and it becomes impossible to relate the conditions to a partial pressure of \ce{H2} since the chemical potential is not an observable of the canonical ensemble.

Second, if we let $\psi = E + \mu_\text{H} N_\text{H}$, where $\mu_\text{H}$ and $N_\text{H}$ are the chemical potential of hydrogen and the number of hydrogen atoms, respectively, we sample the grand canonical ensemble with respect to the H sublattice.
In this case, the chemical potential of hydrogen is specified rather than the concentration, similar to experiment, and it is possible to relate the conditions to a partial pressure of \ce{H2} (see \autoref{sec:mu-to-pressure}).
Moreover, one has access to a free energy derivative since
\begin{equation} \label{eq:sgc-derivative}
    \partial F / \partial c_\text{H} = \mu_\text{H}.
\end{equation}
If we view vacancies as a chemical species, we may also refer to this as sampling in the \gls{sgc} ensemble (with the difference in chemical potential simply being the chemical potential of hydrogen as the chemical potential of vacancies is zero) and this term will be used below.

Finally, we also used the \gls{vcsgc} ensemble \cite{SadErh12, SadErhStu12},
in which $\psi = E + N k_\text{B} T \bar{\kappa} (c + \bar{\phi}/2)^2$, where $\bar{\kappa}$ and $\bar{\phi}$ control the variance and mean of the concentration $c$, respectively.
With this approach we may, unlike the \gls{sgc}, stabilize the system inside multi-phase regions and thereby integrate free energies across phase boundaries using the free energy derivative
\begin{equation}\label{eq:vcsgc-derivative}
    \partial F / \partial c = - 2 N k_\text{B} T \bar{\kappa} \left( \left \langle c \right \rangle + \bar{\phi} / 2 \right),
\end{equation}
where $\left \langle c \right \rangle$ is the average observed concentration.
By equating Eq.~\eqref{eq:sgc-derivative} and \eqref{eq:vcsgc-derivative}, we may also relate an observed state to an external partial pressure of \ce{H2}.

It should be noted that the two sublattices may be sampled in different ensembles in the same \gls{mc} simulation.
In this work, we have used several combinations of the three above-mentioned ensembles as will be detailed below.
For each set of independent parameters, we ran 1,000 \gls{mc} sweeps (1 sweep = $N$ steps, where $N$ is the number of sites in the cell), of which the first 50 were discarded to allow for equilibration.
The simulations were commenced from a random H--vacancy sublattice with equimolar concentrations, whereas the starting configuration of the Pd--Au sublattice was either random, or, in the case of random equilibrium and para-equilibrium (see \autoref{sec:random-para-results}) as well as for the simulation of isotherms, fixed in a predetermined configuration.
The simulations were done with a supercell corresponding to $6 \times 6 \times 6$ times the conventional cell, for a total of 1,728 sites (864 per sublattice).
All \gls{mc} simulations were carried out with the \textsc{mchammer} module of the \textsc{icet} package \cite{AngMunRah19}.

\subsection{Conversion between chemical potential and partial pressure of hydrogen} \label{sec:mu-to-pressure}
Experimentally, the chemical potential of hydrogen, $\mu_{\text{H}} = \frac{1}{2} \mu_{\text{H}_2}$, is controlled via the partial pressure of \ce{H2}, $p_{\text{H}_2}$.
Treating \ce{H2} as an ideal gas, we can write the following relation between its chemical potential and pressure,
\begin{equation} \label{eq:ideal-gas-mu}
    \mu_{\text{H}_2}(T, p_{\text{H}_2}) = \mu^{\circ}_{\text{H}_2}(T) + k_\text{B} T \ln \frac{p_{\text{H}_2}}{p^{\circ}_{\text{H}_2}}.
\end{equation}
Here, $p^{\circ}_{\text{H}_2}$ is a reference pressure and $\mu^{\circ}_{\text{H}_2}(T)$ is the temperature dependent chemical potential of \ce{H2} at this reference pressure.
The latter can in principle be evaluated with a combination of \gls{dft} calculations and thermochemical tables, but since small errors in $\mu^{\circ}_{\text{H}_2}(T)$ have a large impact on $p_{\text{H}_2}$, we chose to make a two-parameter fit.
Specifically, we extracted experimental data of the partial pressures of the $\alpha \rightarrow \beta$ transition in pure Pd at different temperatures, and fitted a linear function to these $(\ln(p), T)$ data points.
We then fitted a linear function to the points $(2 \mu_\text{H}, T)$ at which we observed the phase transition in our own simulations.
The conversion is then given by
\begin{equation}
    p_{\text{H}_2} = \exp\left( \frac{2 \mu_\text{H} + \Delta P(T)}{k_\text{B}T} \right)
\end{equation}
where $\Delta P(T)$ is the difference between the two linear fits, specifically $\Delta P(T) = \unit[1.2]{meV/K} \cdot T - \unit[0.3547]{eV}$ (for more details, see \autoref{sfig:mu-to-pressure}).
It should be noted that breakdown of the ideal gas approximation at very high \ce{H2} pressures will make our conversion between chemical potential and pressure increasingly inaccurate.

\subsection{Short and long range order quantification}
We quantify \gls{sro} with the Warren--Cowley order parameter,
\begin{equation}
    \alpha_\text{A--B} = 1 - \frac{n_\text{A--B}}{Z c_\text{A} c_\text{B}},
\end{equation}
where A and B refer to the two species involved (Au and Pd or H and vacancy, depending on sublattice), $n_\text{A--B}$ is the total number of unlike bonds $A$--$B$ in a certain shell (such as nearest neighbors), $Z$ is the total number of bonds (regardless of decoration), and $c_\text{A}$ and $c_\text{B}$ is the overall concentration of A and B, respectively.
To quantify \gls{sro} between the two sublattices, we generalize the above expression,
\begin{equation} \label{eq:sro}
    \alpha_\text{Au/Pd--H/vac} = 1 - \frac{n_\text{Pd--H,Au--vac}} {Z (c_\text{H} c_\text{Pd} + c_\text{Au} c_\text{vac})},
\end{equation}
where $n_\text{Pd--H,Au--vac}$ is the total number of pairs that are either Pd--H or Au--vacancy.

To quantify \gls{lro} on the Pd--Au sublattice, we use a standard expression for the structure factor,
\begin{equation} \label{eq:lro}
    S(\mathbf{q}) = \frac{1}{N^2}\sum_{ij} f_i f_j e^{-i \mathbf{q} \cdot (\mathbf{R}_i - \mathbf{R}_j)},
\end{equation}
where the sum runs over all pairs $i,j$ of sites in the sublattice (with a total of $N$ sites), $\mathbf{R}_i$ is the position of atom $i$, and $f_i = 1$ if atom $i$ is Au and $-1$ if it is Pd.

\subsection{Phase diagram construction}
\label{sec:phase-diagram-construction}
Once mixing free energy curves $F_\text{mix}(c_\text{H}, c_\text{Au}, T)$ have been calculated from \gls{mc} simulations, phase diagrams can be constructed by identifying the regions where the free energy curve lies above its convex hull.
(This construction relies on the assumption that the interface between the two phases is incoherent; hysteresis stemming from coherency will in general increase the apparent solubility \cite{SchKha06, SchKhaCar20}.)
The boundaries of these regions constitute the phase boundary, and to obtain a continuous phase boundary, we fitted, primarily as a guide to the eye, the observed boundary points to the function
\begin{equation}
    f(c) = a \frac{2c - 1}{\ln c - \ln(1 - c)} + L(c)
\end{equation}
where $a$ is a fitting parameter and $L(c)$ is a fourth-order Redlich--Kister polynomial with five fitting parameters.

The phase diagram at full equilibrium is more intricate to construct since it involves integration of data with noise over a path in two dimensions.
To this end, we averaged the integration over multiple paths and constructed the phase diagram based on several convex hull constructions.
For details, see \autoref{snote:phase-diagram-full-equilibrium}.

\section{Results and discussion}

\subsection{Cluster expansion construction}

A \gls{ce} operates under the assumption that the atoms reside on a fixed lattice.
In practice the effect of relaxation from the ideal lattice sites is, however, incorporated effectively through the fitting of the \glspl{eci} $J_\alpha$.
For this treatment of relaxation to be feasible, the relaxed structure must not deviate too strongly from the ideal lattice \cite{NguRosRee17}.
An analysis of our \gls{dft} calculations revealed that in many structures, in particular those with a high Au content, relaxations from the ideal lattice sites were large, sometimes more than half the distance to the nearest neighbor on the other sublattice (\autoref{sfig:relaxations}a).
To avoid training the \gls{ce} on data that is not well represented by a lattice model, we excluded all structure in which at least one atom had relaxed more than \unit[0.5]{\AA} from its ideal site, which is equivalent to approximately 25\% of the distance to its nearest neighbor.
Furthermore, the relaxed cell shape sometimes deviated from the cubic cell.
In particular, pure AuH with 100\% H is mechanically unstable and relaxes into a monoclinic structure (it should be noted that pure Au hydrides are considered unstable even under very high \ce{H2} pressures \cite{RahHofAsh17, DonSchGre13}).
To further investigate the crystal structure, we therefore also analyzed the shearing of the cell.
Here, we define the shear strain tensor $\Delta A$ by subtracting the volumetric strain from the Biot strain tensor \cite{Bio65} (defined as $\mathbf{U}-\mathbf{I}$, where $\mathbf{U}$ is in turn defined by the polar decomposition of the deformation tensor, $\mathbf{F}=\mathbf{U}\mathbf{P}$), and a scalar measure of the strain is obtained by taking the Frobenius norm of $\Delta A$.
On average, higher concentrations of Au and H tend to increase the shear strain (\autoref{sfig:relaxations}b).

Since structures with a very high Au content are not experimentally relevant except under fairly exotic conditions, we chose to exclude all structures with Au concentration above 50\%.
With the limitation to structures with no displacements larger than \unit[0.5]{\AA} and no Au content above 50\%, all structures with a shear strain $||\Delta A||_\text{F} > 0.4$ were effectively excluded.
In the end, we were left with 1,305 structures, out of which 1,255 were used for training the \gls{ce}, with the remaining 50 used as test set to evaluate the performance of the final model.

\begin{figure*}
    \centering
    \includegraphics{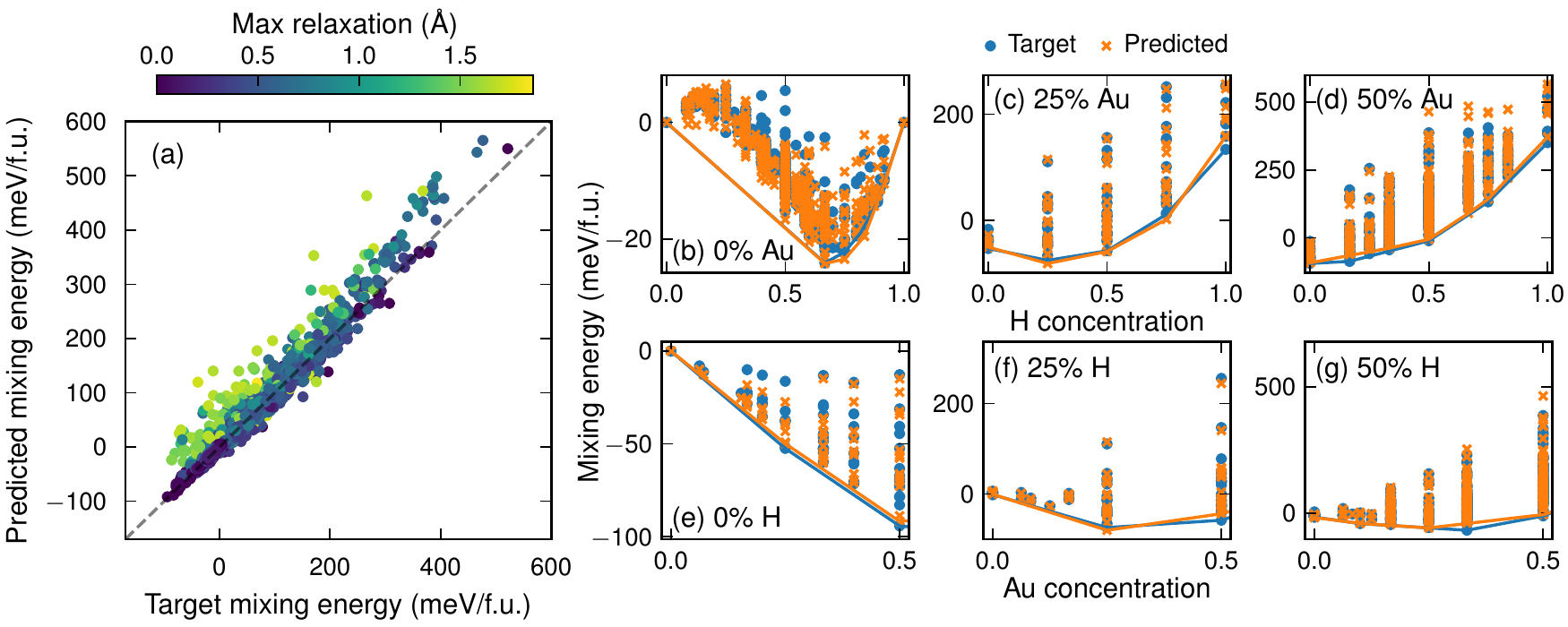}
    \caption{Mixing energies as predicted with the CE developed in this work compared to target energies from DFT calculations.
    Structures for which the largest relaxation distance was small (dark blue dots in (a)) are generally well reproduced by the CE, whereas the energy of structures with large relaxation distances are overestimated by the CE (yellow dots in (a)).
    Only structures with maximum relaxation distance less than \unit[0.5]{\AA} were included when training the CE.
    Mixing energies as a function of H concentration at fixed Au concentration (b--d) and as a function of Au concentration at fixed H concentration (e--g) reveal that the general energy landscape is well reproduced in the concentration intervals considered here.
    }
    \label{fig:ce-quality}
\end{figure*}

The mixing energies span an energy range of approximately $\unit[-100]{meV/f.u.}$ to $\unit[500]{meV/f.u.}$ (\autoref{fig:ce-quality}a).
This wide range of energies is a challenge for the \gls{ce} model, since some of the important phenomena occur on a much more narrow energy scale.
Consider, for example, the two-phase region of pure Pd hydride.
This two-phase region is the result of a concave mixing energy curve between $c_\text{H}=0$ and $0.67$ (\autoref{fig:ce-quality}b),
and the distance between the lowest energy structures and the convex hull is at most a few \unit{meV/\gls{fu}} (with the limited cell size in our training data).
Meanwhile, if we increase the Au content, relevant variations occur on a scale that is at least one order of magnitude larger.
To ensure that the important variations at low Au content are properly captured, we multiplied the cluster vectors and target energies with a weight $w(c_\text{Au}) = 1 + 10(0.5- \text{c}_\text{Au})$ prior to fitting, chosen to make the variations in mixing energy occur within the same order of magnitude over the full range of Au concentrations.
We note that non-uniform weighting of structures as well as the imposing of constraints to exactly reproduce some structures is done at the cost of a larger error for other structures, and inevitably a larger root mean square error over the full test set.
Yet, it is more important to capture the relevant aspects of the energy landscape properly rather than having the lowest possible root mean square error.

The \gls{ce} constructed in this fashion yields a root mean square error for the mixing energies of \unit[8.3]{meV/\gls{fu}} over the training set  and of \unit[15.3]{meV/\gls{fu}} over the test set.
To put these number in perspective, it is instructive to compare them to the wide span of mixing energies in this system (\autoref{fig:ce-quality}).
The largest errors are frequently found among structures with a large content of Au, which partially explains the large difference in error between training and test set, as the latter contained a higher proportion of structures with high Au content.
The predicted energy of all structures (including structures that relaxed too far relative to the reference structure (see \autoref{sfig:relaxations}) but excluding those with $c_\text{Au} > 0.5$) has a good albeit not perfect correlation with the corresponding \gls{dft} energies (\autoref{fig:ce-quality}a, $R^2=0.934$).
The energies of structures with long relaxation distances are sometimes severely overestimated.
This is unsurprising since they were not included in the training data and cannot be expected to be well described by the \gls{ce} in any case.
Many of the poorly described structures are those with higher mixing energies in a series of configurations with the same composition (\autoref{fig:ce-quality}b--g), and hence are in general less relevant for the thermodynamics of the system.

\begin{figure}
    \centering
    \includegraphics{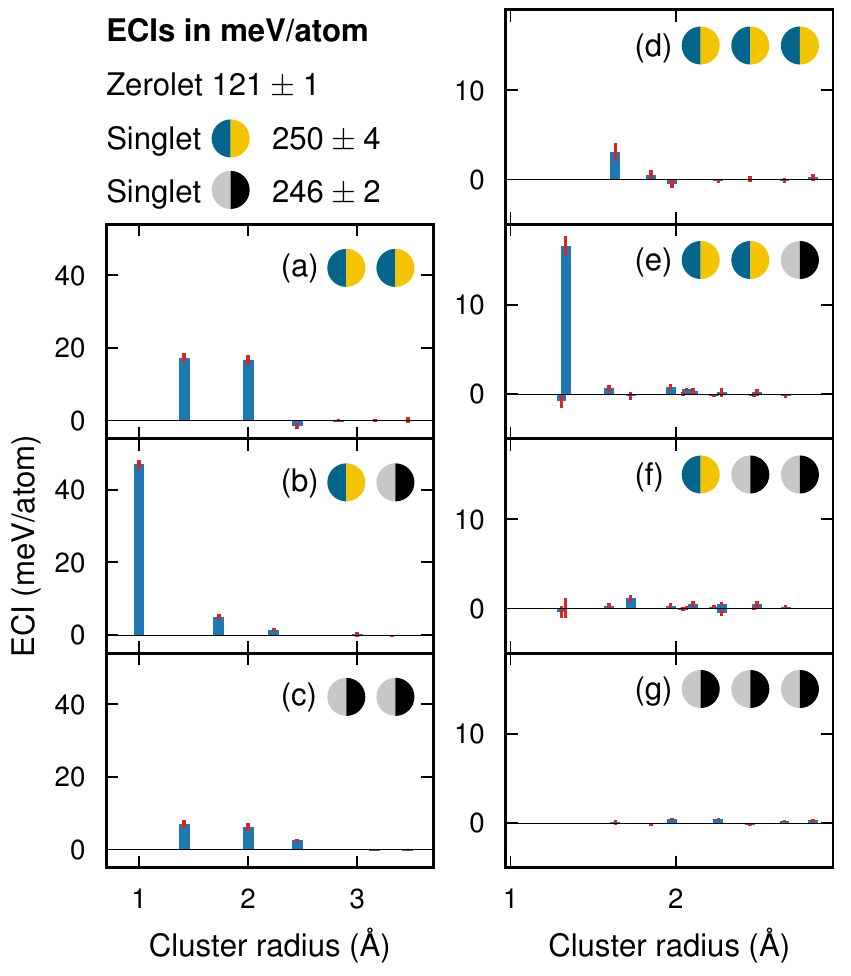}
    \caption{
        Effective cluster interactions (ECIs) in the cluster expansion developed in this study, as function of the radius of the cluster, here defined as the average distance between the atoms in the cluster and their center of mass (assuming all atoms have the same mass).
        Pair ECIs are split in three categories, interactions between Au/Pd and Au/Pd (a), between Au/Pd and H/vacancy (b), and between H/vacancy and H/vacancy (c).
        In the same vein, triplet interactions are split into four categories depending on the sublattices involved (d)--(g).
        Error bars, defined as two standard deviations over a set of 100 randomly sampled cluster expansions, are shown in red at the tip of each bar.
    }
    \label{fig:ecis}
\end{figure}

The \glspl{eci} of the constructed \gls{ce} exhibit physically sound behavior in the sense that the strongest interactions are those that occur at a short distance, and triplet interactions are generally smaller than pair interactions (\autoref{fig:ecis}).
Based on the \glspl{eci}, we can already at this point anticipate roughly how the system will behave.
The strongest interaction is the one between Au/Pd and its nearest neighbor H/vacancy.
We can anticipate that H will be attracted to Pd and vacancies to Au.
H--H interactions, on the other hand, are comparatively weak, whereas the positive, short-ranged Au/Pd--Au/Pd \glspl{eci} indicate that the formation of Pd--Au pairs is energetically favorable.

\subsection{Ordered phases in hydrogen-free Pd--Au}
\begin{figure}
    \centering
    \includegraphics[scale=0.42]{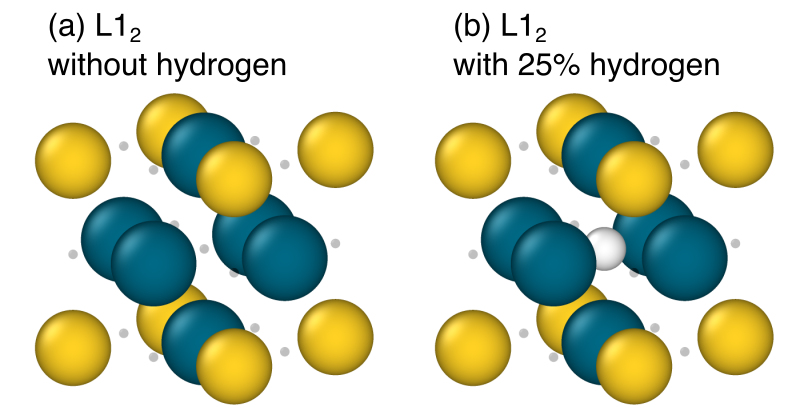}
    \caption{The L1$_2$ phase without hydrogen (a) and with 25\% hydrogen (b).
    Palladium, gold, and hydrogen are shown in blue, yellow, and white, respectively, whereas vacancies are shown as grey dots.
    In this ordered phase, hydrogen preferentially occupies the site surrounded by nearest neighbor Pd atoms.
    Note that this choice of cell does not give an accurate representation of compositions; the structures have (a) 25\% Au and 75\% Pd and (b) the same composition on the Pd--Au sublattice and 25\% hydrogen and 75\% vacancies on the other.
    }
    \label{fig:L12-phases}
\end{figure}

Although the focus of our work is on the thermodynamics of hydrogen in Pd--Au, we begin our analysis with the Pd--Au alloy in absence of hydrogen as ordering turns out to be also relevant with regard to the hydrogenated system, as discussed below.
As our \gls{ce} is only fitted to structures with 50\% Au or less, we restrict our analysis to this composition interval.
The mixing energies of structures with 12 Pd/Au atoms or less are negative, meaning mixing is favorable.
It is, however, still possible that phase separation occurs between ordered (intermetallic) phases.
\Gls{lro} in Pd--Au has been reported in experiments on thin films \cite{MatNagKak66, KawInoOga71} and nanoparticles \cite{NelNugAll14}, but has generally not been found in the bulk, suggesting that the energetics causing ordering is either enhanced by surface stress or has critical temperatures too low for ordering to occur readily in bulk samples.
By enumerating structures with up to 12 atoms in the cell, our \gls{ce} identifies three structures on the convex hull between 0 and 50\% Au: the trivial case of pure Pd, \ce{AuPd3} in the L1$_2$ configuration (\autoref{fig:L12-phases}a) and AuPd in the L1$_0$ configuration.
These findings are largely consistent with other computational studies, although other structures have occasionally been found on the convex hull \cite{BarBluMul06, BerLeg20}.

To investigate at below which temperature the L1$_2$ phase will appear, we ran \gls{mc} simulations in the canonical ensemble, i.e., with the composition fixed at 25\% Au in Pd, and tracked the \gls{lro} parameter (see \autoref{eq:lro}) as a function of temperature.
The results revealed a critical temperatures at \unit[180]{K} (\autoref{sfig:lro-L12}), with a span among the sampled \glspl{ce} from less than \unit[150]{K} to \unit[260]{K}.

\subsection{Hydrogen in pure Pd}
\begin{figure}
    \centering
    \includegraphics{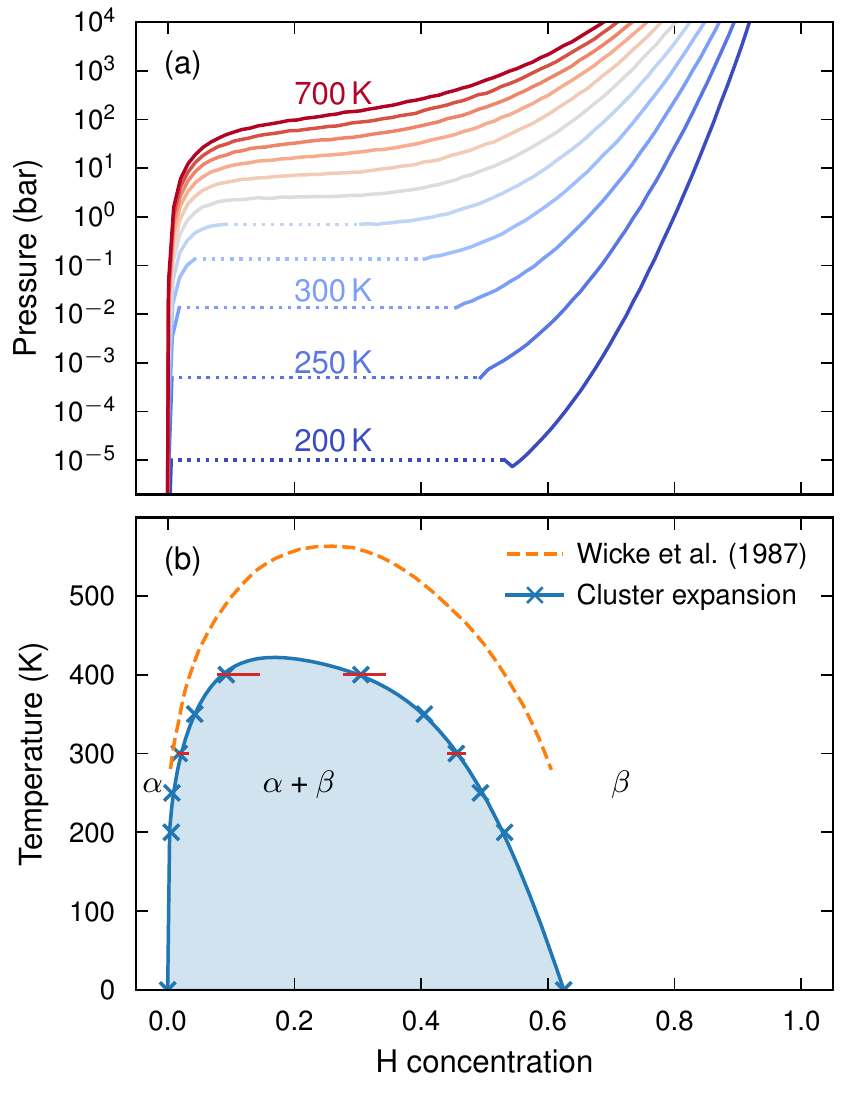}
    \caption{Pressure--composition isotherms in steps of \unit[50]{K} for pure Pd hydrides (a).
    Phase transitions from the hydrogen-poor $\alpha$ phase to the hydrogen-rich $\beta$ phase have been drawn with a dashed line.
    These dashed lines indicate the phase boundaries of the two-phase region (blue crosses in (b), blue line is a fit to guide the eye).
    Compared to experimental results \cite{WicBla87} (orange line in b), the critical temperature is underestimated, but the uncertainty from the choice of CE is significant, as indicated by the spread of the miscibility gap at 300 and \unit[400]{K} from ten sampled CEs (red lines indicate difference between minimum and maximum value predicted by these ten CEs).
    Here, we have disregarded ordered phases with hydrogen concentrations above 0.67, all of which have critical temperatures below \unit[200]{K}.
    }
    \label{fig:purePd-thermo}
\end{figure}

We are now ready to add hydrogen to the system.
Since the pure Pd hydride (no Au) has been more thoroughly studied than the alloy hydride, we begin in this limit.
We carried out \gls{vcsgc}-\gls{mc} simulations on the H/vacancy lattice with temperatures between \unit[200]{K} and \unit[700]{K} in steps of \unit[50]{K}.
The thus obtained isotherms (\autoref{fig:purePd-thermo}a) indicate a low solubility of H in Pd until a ``threshold pressure'' has been reached.
For sufficiently low temperatures, there is then a phase transition from a hydrogen-poor $\alpha$ phase to a hydrogen-rich $\beta$ phase, which can be detected by integrating the free energy derivative and identify the regions where the free energy curve lies above its convex hull (as outlined in \autoref{sec:phase-diagram-construction}).
The corresponding phase diagram is largely consistent with the experimental phase diagram constructed by Wicke and Blaurock \cite{WicBla87} (\autoref{fig:purePd-thermo}b).
The critical temperature, which is challenging to estimate accurately both experimentally and by means of \gls{mc} simulations, differs approximately \unit[140]{K} (\unit[560]{K} experimentally and \unit[420]{K} based on our \gls{ce}).
Although this is a fairly large discrepancy, it should be noted that small variations in the \glspl{eci} have a large impact on the critical temperature; at temperatures close to the critical temperature, uncertainties in the phase boundary become large (quantified by ten sampled \glspl{ce}; the spread from these ten \glspl{ce} is indicated with red horizontal bars in \autoref{fig:purePd-thermo}b).
Discrepancies between simulation and experiment are thus not unexpected, especially given that it is also challenging to construct the phase diagram experimentally.

\subsection{Hydrogen loading of fixed Pd--Au lattice: random and para-equilibrium} \label{sec:random-para-results}
We are now ready to include Au in the \gls{mc} simulations.
To this end, we note that the two sublattices are markedly different from a kinetic perspective.
Hydrogen diffuses easily even at low temperatures, while Pd and Au diffuse over a much longer timescale.
To reach full equilibrium, in which the Pd and Au atoms rearrange in response to a change in hydrogen environment, a very slow experiment with \ce{H2} pressure kept constant over extended periods of time, is required.
If the \ce{H2} pressure is more quickly increased/decreased, we may instead assume that the Pd--Au sublattice is frozen and that a metastable equilibrium is reached by rearrangement of the hydrogen atoms only.
The chemical order on the Pd--Au sublattice would then typically be dictated by the conditions during fabrication of the alloy, which usually involve a form of thermal annealing.
Here, we distinguish two extremes of such a metastable equilibrium: \emph{para-equilibrium}, in which the Pd--Au ordering is equilibrated at \unit[300]{K} in the absence of \ce{H2}, and \emph{random equilibrium}, in which the Pd--Au lattice is randomized, corresponding to quenching of the system after equilibration at a very high temperature in absence of \ce{H2}.

We can model these situations by only carrying out \gls{vcsgc}-\gls{mc} flips on the H/vacancy sublattice.
For the random equilibrium, we commence from a simulation cell in which the Pd and Au atoms have been randomly distributed, whereas for para-equilibrium, we run a canonical \gls{mc} simulation with only Pd and Au and randomly pick a configuration as the fixed Pd--Au lattice.
To suppress spurious effects from the particular choice of Pd--Au lattice, we averaged our results over five instances of both cases.

\begin{figure}
    \centering
    \includegraphics{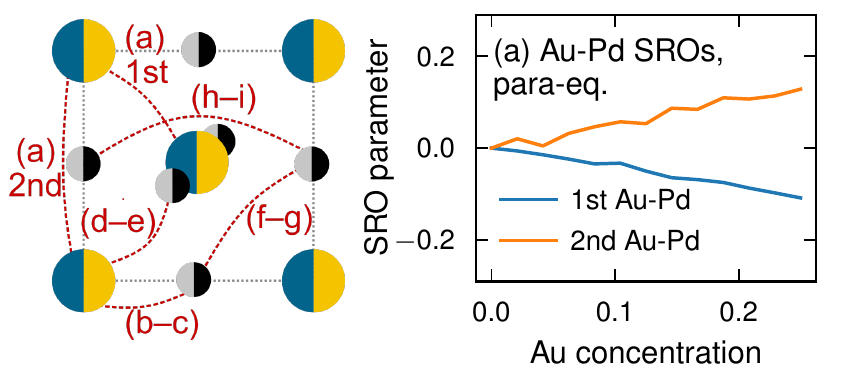}
    \includegraphics{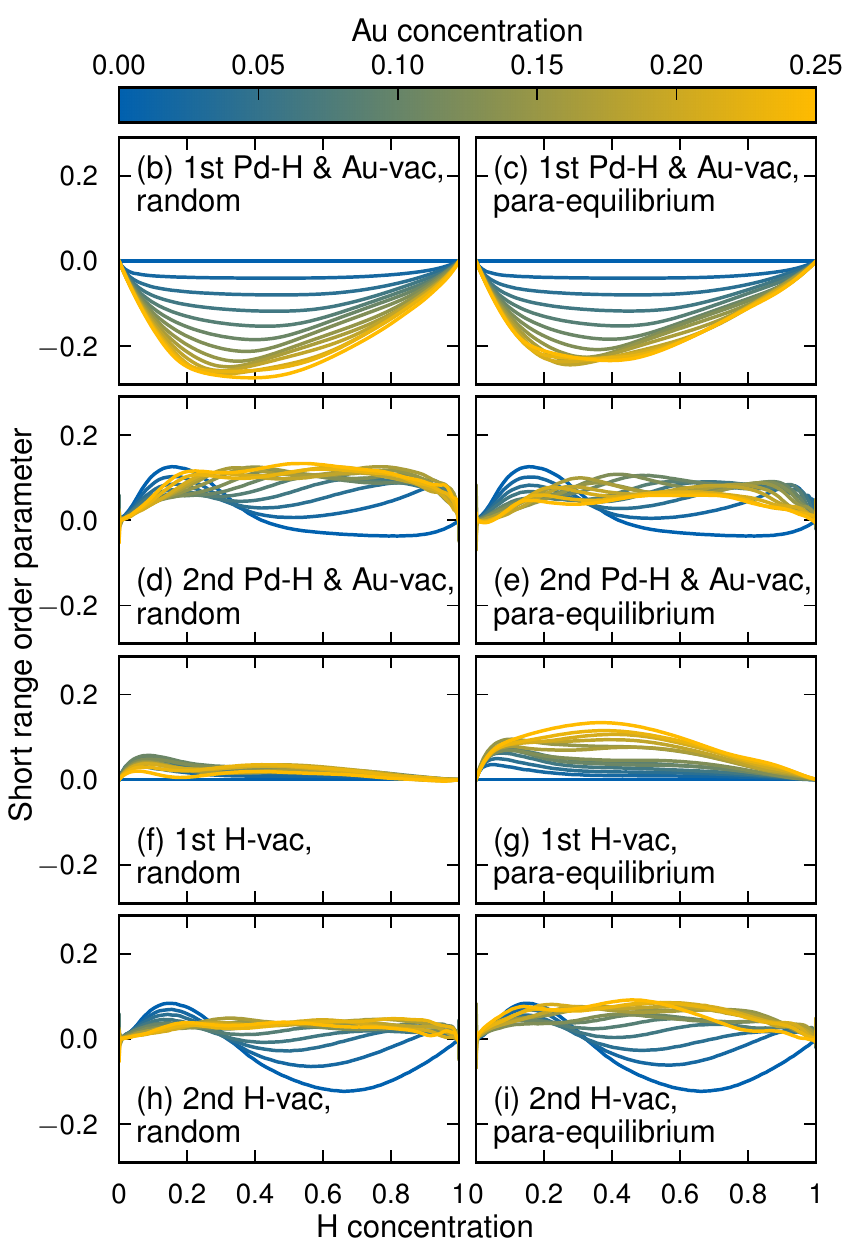}
    \caption{\Gls{sro} at \unit[300]{K} in Pd--Au hydride with the Pd--Au lattice fixed in a random configuration (left column) and as equilibrated in absence of H (para-equilibrium, right column).
    The \gls{sro} in the nearest-neighbor shell between the Pd/Au and the H/vacancy lattices (a--b) is negative, which, with our definitions (see Eq.~\eqref{eq:sro}), means that H tends to sit close to Pd and vacancies close to Au.
    In the second Au/Pd shell (c--d), a weak reverse trend is seen throughout most of the H concentration range.
    In the first H/vacancy shell (f--g), H--H pairs are favored, only weakly in random equilibrium but stronger in para-equilibrium.
    The behavior is more complex in the second H/vacancy shell, with ordering tendencies being dependent on both H and Au concentrations.
    }
    \label{fig:sro-rand-and-para}
\end{figure}

These \gls{mc} simulations reveal weak \gls{sro} (\autoref{fig:sro-rand-and-para}).
We first note that the equilibration of the Pd--Au lattice in absence of hydrogen yields a weak propensity for the system to form unlike (Pd--Au) nearest neighbor bonds, whereas bonds that are alike (Au--Au and Pd--Pd) are weakly favored among next-nearest neighbors (\autoref{fig:sro-rand-and-para}a; in random Au--Pd the corresponding order parameters are zero by construction).
For nearest neighbor Pd/Au--H/vacancy pairs (\autoref{fig:sro-rand-and-para}b--c), the \gls{sro} parameter is negative, indicating that H is more likely to occupy sites next to Pd than next to Au.
This confirms what has been observed both with Mössbauer spectroscopy \cite{WagKarPro83} and first-principles calculations \cite{MamZhd20}, namely that it is energetically unfavorable for hydrogen to occupy the sites closest to a Au atom.
For next-nearest Pd/Au--H/vacancy neighbors (\autoref{fig:sro-rand-and-para}d--e), on the other hand, the effect is reversed, albeit weaker.
The change of sign between first and second nearest-neighbor was observed also by Sonwane \textit{et al.} \cite{SonWilMa06} in a model based on \gls{dft} calculations.
It is furthermore consistent with the common argument that chemistry causes repulsion between neighboring Au and H, whereas dilation of the lattice by a Au atom gives rise to a more long-ranged elastic attraction \cite{Fuk06}.

The most pronounced difference between the two types of equilibria is found for the most short-ranged H/vacancy pair (\autoref{fig:sro-rand-and-para}f--g), for which the system in para-equilibrium has a stronger tendency to form H--H pairs.
For next-nearest neighbor H/vacancy pairs (\autoref{fig:sro-rand-and-para}h--i) a more complex behavior emerges, with the sign of the \gls{sro} parameter being dependent on both H and Au concentration.
For high H and Pd content, the \gls{sro} parameter is negative, indicating that H atoms are distributed throughout the material, as H--H pairs are not favored.
With a high content of Au and/or low content of H, the opposite behavior occurs, i.e., there is a slight excess of H--H pairs.
This effect is somewhat more pronounced in para-equilibrium.

\begin{figure}
    \centering
    \includegraphics{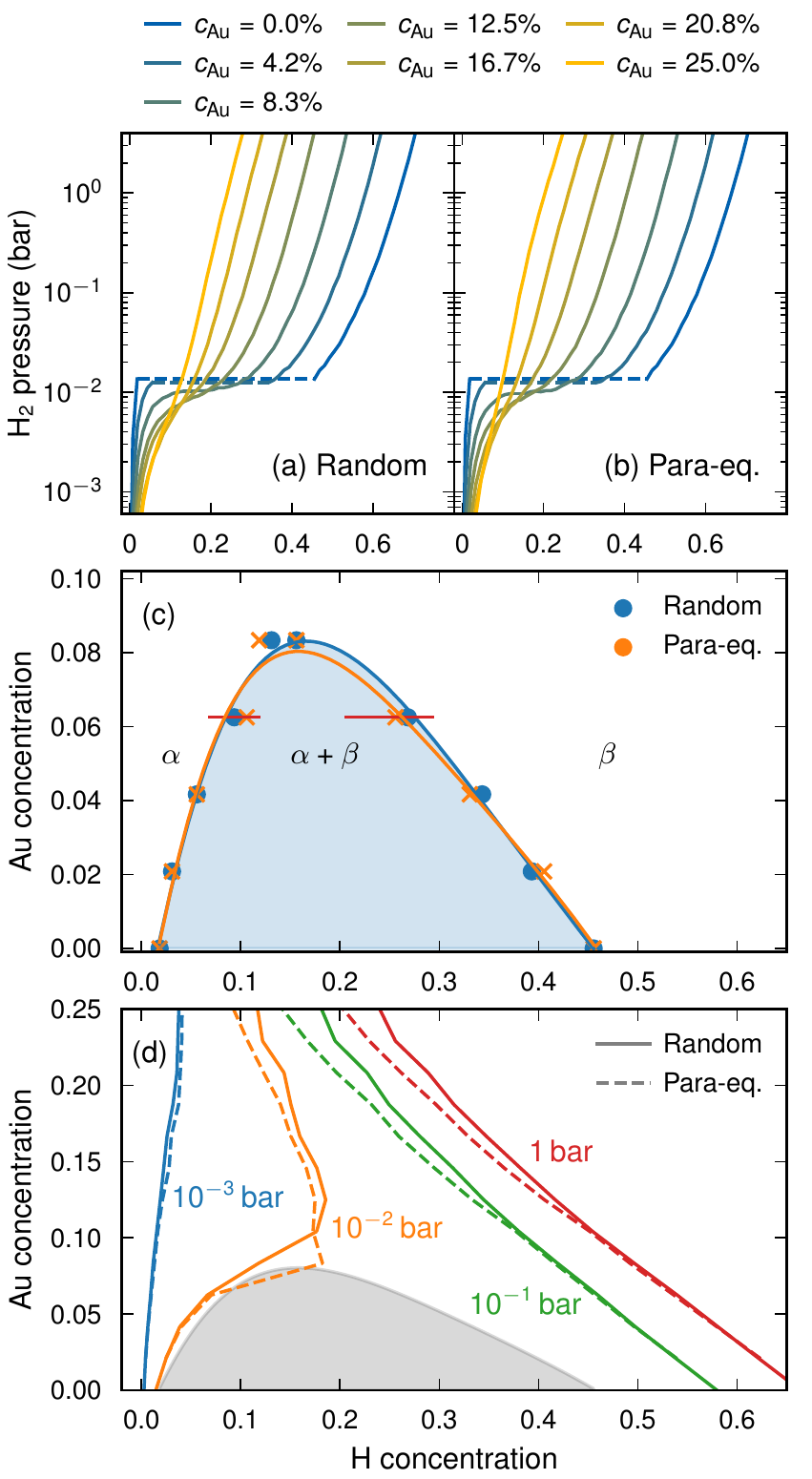}
    \caption{
        Pressure--composition isotherms at \unit[300]{K} for Pd--Au hydride with (a) a random Pd--Au sublattice and (b) in para-equilibrium, i.e., with the Pd--Au lattice equilibrated at \unit[300]{K} in absence of H.
        Phase transitions from the H-poor $\alpha$ phase to the H-rich $\beta$ phase are indicated with a dashed line.
        The corresponding two phase diagrams (c) are almost identical, with the two-phase region closing at approximately 8-9\% Au.
        Red lines at 6.3\% Au indicate the range of values obtained from ten sampled CEs.
        By studying the hydrogen uptake at fixed \ce{H2} pressure as a function of Au concentration (d), it is clear that differences between random and para-equilibrium are found primarily at high \ce{H2} pressures, where at the same pressure the random system absorbs a slightly larger amount of hydrogen than the system in para-equilibrium.
        Larger variations are also seen at $\unit[10^{-2}]{bar}$ (orange line), since this it close to the plateau pressure.
        The phase diagram in para-equilibrium is indicated with transparent grey in (d).
    }
    \label{fig:para-and-rand-thermo}
\end{figure}

Given the quantitative difference in chemical ordering between random Au--Pd and para-equilibrium, we may now look for consequences for their respective thermodynamics.
It turns out, however, that their \ce{H2} pressure--composition isotherms (\autoref{fig:para-and-rand-thermo}a--b) are very similar.
It is only at fairly high Au concentrations, approaching 25\% Au, that the isotherms of random and para-equilibrium exhibit discernible differences.
At high Au concentrations and high \ce{H2} pressures, the hydrogen uptake is slightly higher in random equilibrium than in para-equilibrium (\autoref{fig:para-and-rand-thermo}d).
Although this difference is small, there seems to be a weak but consistent trend; when \gls{sro} emerges due to lower annealing temperature, the hydrogen uptake at \unit[300]{K} goes down at pressures above approximately \unit[10]{mbar} (\autoref{sfig:uptake-anneal-temp}).
The difference seems to be the largest at a hydrogen pressure around \unit[1]{bar}, where the fully random alloy with 25\% Au absorbs almost 4 percentage points more hydrogen than the one equilibrated in \unit[300]{K}.
These results seem to be in agreement with Chandrasekhar and Sholl \cite{ChaSho14}.
To summarize, the \gls{sro} that emerges at low temperatures in Pd--Au makes the material absorb slightly less hydrogen at high \ce{H2} pressures.

The discontinuity in the Au-poor isotherms, which is the hallmark of the phase transition from $\alpha$ to $\beta$, disappears quickly when the Au content is increased and it does so in almost exactly the same way for random and para-equilibrium.
Consequently, the respective phase diagrams are essentially identical (\autoref{fig:para-and-rand-thermo}c).
Our results predict a critical Au concentration for the $\alpha + \beta$ two-phase region around 8--9\%.
This is significantly lower than experimental measurements, which yield estimates that vary from 10--15\% (at \unit[303]{K}) \cite{LuoWanFla10} to around 17\% (at \unit[298]{K}) \cite{MaeFla65}.
The prediction of the \textit{solvus} line is, however, sensitive to small variations in the \glspl{eci} (quantified by the spread obtained by sampling ten \glspl{ce} and indicated by red horizontal bars in \autoref{fig:para-and-rand-thermo}c).
Thus both our prediction and the experimental value are associated with significant uncertainties.

\subsection{Hydrogen loading in full equilibrium}
\begin{figure}
    \centering
    \includegraphics{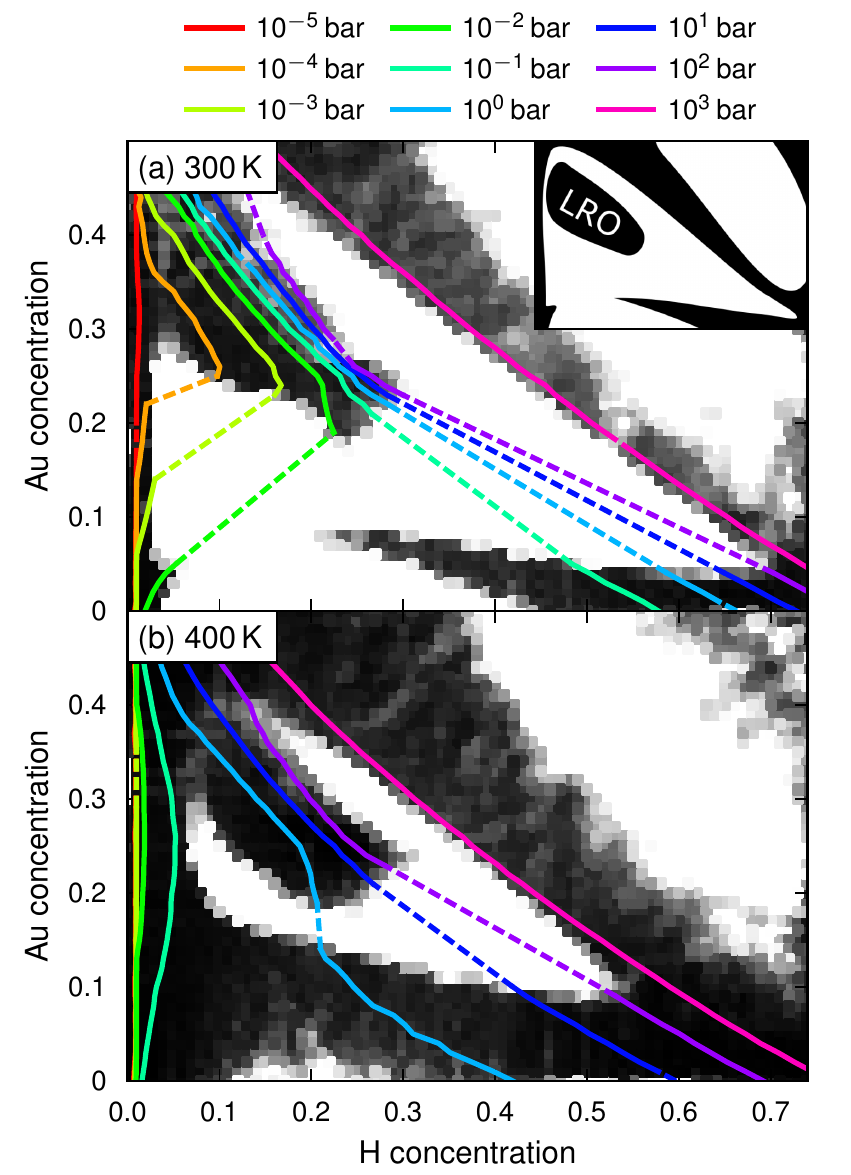}
    \caption{Phase diagram of hydrogenated Pd--Au in full equilibrium at (a) \unit[300]{K} and (b) \unit[400]{K}.
    Black areas indicate single-phase regions, whereas white areas are multi-phase regions.
    Isobars (lines of fixed \ce{H2} partial pressure) are drawn with colored lines.
    Many of the isobars converge in a region around 25\% Au and 25\% H, where the system exhibits \gls{lro}.
    The remaining dark areas represent a solid solution.
    The inset shows a schematic representation of the phase diagram to aid its interpretation.
    }
    \label{fig:full-phase-diagram-T300-T400}
\end{figure}

We may now ask what will happen if we allow the Pd--Au lattice to rearrange as we expose it to hydrogen.
This situation is commonly referred to as full (or complete) equilibrium.
It should be noted that this is an idealization that is very time-consuming to achieve in practice, as in most experiments the \ce{H2} pressure is not maintained long enough to allow for Pd and Au to diffuse to a sufficient extent.
Yet, it is important as it provides the thermodynamic driving force for the changes that do occur, although they may only rarely take the system to full equilibrium.
To investigate the full equilibrium, we carried out \gls{mc} simulations in the \gls{vcsgc} ensemble on both sublattices.
Subsequently we obtained the free energy by integration across the concentration plane using many different integration paths, and sampled the convex hull based on these different integration paths (for details, see \autoref{snote:phase-diagram-full-equilibrium}).
Thereby we obtained a heat map of the probability that certain compositions are on the convex hull, which can be interpreted as (an isothermal cut of) the phase diagram (\autoref{fig:full-phase-diagram-T300-T400}; see \autoref{sfig:full-phase-diagram-T250-T500-6x6x6} for data at 250 and \unit[500]{K}).
The two-phase region at low Au content observed in para and random equilibrium is present in the full phase diagram at \unit[300]{K} as well, but is accompanied by a much larger multi-phase region at higher Au concentrations (\autoref{fig:full-phase-diagram-T300-T400}a).
At \unit[400]{K}, the former two-phase region is almost gone, but the multi-phase region at higher Au contents is still present.
This new multi-phase region that appears in the full phase diagram, as opposed to random and para-equilibrium, is largely driven by a particularly stable composition interval around 25\% Au and approximately 10--30\% H, where at \unit[300]{K} isobars ranging from $\unit[10^{-3}]{bar}$ to $\unit[10^2]{bar}$ all converge.
A closer look at this composition interval reveals that \gls{lro} develops here (\autoref{sfig:lro-maps}).
Specifically, the Pd--Au sublattice orders in the L1$_2$ phase, in which the atoms are arranged such that 25\% of the hydrogen sites are octahedral ``cages'' with only Pd nearest neighbors (\autoref{fig:L12-phases}b).
As has been previously reported \cite{MamZhd20} and can be expected from the \glspl{eci} (\autoref{fig:ecis}), occupation of hydrogen at such sites is energetically particularly favorable, and renders this long-range ordered phase particularly stable.
At \unit[500]{K} (\autoref{sfig:full-phase-diagram-T250-T500-6x6x6}b), \gls{lro} no longer forms and the phase diagram is largely featureless, i.e., a solid solution can be expected at any concentration (except possibly at high Au content and extremely high \ce{H2} pressures).

The phase diagram in \autoref{fig:full-phase-diagram-T300-T400}a thus predicts that if a Pd--Au alloy at \unit[300]{K} with, say, approximately 15\% Au is subjected to a \ce{H2} pressure between approximately $\unit[10^{-3}]{bar}$ and $\unit[10^2]{bar}$, phase separation will occur.
Of the resulting two phases, one will be the L1$_2$ phase with between 10 and 30\% H.
The character of the other phase depends on the \ce{H2} pressure; if the pressure is below $\unit[10^{-2}]{bar}$, the second phase will be the dilute $\alpha$ phase (with Au content strongly dependent on \ce{H2} pressure) whereas if the pressure is above approximately $\unit[10^{-2}]{bar}$, the second phase will be the hydrogen-rich $\beta$ phase with approximately 5\% Au.

The existence of a long-range ordered phase in hydrogen-treated Pd--Au has been reported experimentally by Lee \textit{et al.} \cite{LeeNohFla07}, who observed a super-lattice phase in Pd--Au with 19\% Au.
The structure of this phase could not be determined but the authors reported it to be more complex than L1$_2$.
It should perhaps not be ruled out, however, that the off-stoichiometric composition as well as defects such as anti-phase boundaries, which occurred frequently in some of our simulations, may have played a role in these experiments.

Another region of interest in \autoref{fig:full-phase-diagram-T300-T400} is the single-phase region that the $\unit[10^3]{bar}$ isobar traces out (pink line in \autoref{fig:full-phase-diagram-T300-T400}).
Here, the \gls{sro} parameter for nearest neighbor Pd--Au pairs switches sign, indicating a transition from an excess of Pd--Au pairs to an excess of Au--Au and Pd--Pd pairs (\autoref{sfig:vcsgcb-sro}).
Inspection of the \gls{mc} trajectories at these composition reveals that the system has started to cluster into hydrogen-poor Au clusters and hydrogen-rich Pd clusters, both only a few atoms large.
Our model consider this phase thermodynamically stable and \emph{not} merely a first step towards full phase separation.
It seems plausible that a phase like this may provide a favorable balance between the short-ranged repulsion between Au and H and the long-ranged elastic attraction caused by dilation of the Pd lattice by adjacent Au clusters.
It is not surprising that this phase has not been reported experimentally given the very high \ce{H2} pressure required for it to form ($\gtrsim \unit[10^3]{bar}$).
It should be noted that the pressure specified in this region is highly approximate, as this is outside the applicability range of the ideal gas law.

\subsection{Impact of annealing conditions on hydrogen solubility}
Although full equilibrium will usually not be reached in the time frame of a typical hydrogen loading experiment, there may be circumstances in which full equilibrium is approached.
For example, annealing of Pd--Au is sometimes done in the presence of hydrogen in order to prevent oxidation.
Although the pressure of \ce{H2} is then typically too low to have a significant impact, the phase diagram \autoref{fig:full-phase-diagram-T300-T400} shows that complex behavior may emerge in the presence of hydrogen, especially if the temperature is not too high.
We may consider a fairly typical situation in which the alloy is annealed at a particular temperature and \ce{H2} pressure, after which hydrogen absorption/desorption isotherms are measured at another temperature.
Since the latter absorption/desorption is usually carried out at a much lower temperature than the annealing, and during a much shorter period of time, it is reasonable to assume that the Pd--Au sublattice gets frozen in during annealing and does not change when measuring the isotherm.
The chemical ordering on the Pd--Au sublattice would then be determined entirely by the conditions during annealing.

We now mimic such an experiment by simulating isotherms at \unit[300]{K} in Pd--Au with 25\% of Au annealed in different conditions.
To this end, we first run \gls{mc} simulations at a specified annealing temperature with canonical \gls{mc} swaps on the Pd--Au sublattice, and \gls{sgc} \gls{mc} flips on the H--vacancy sublattice, using a chemical potential corresponding to a fixed \ce{H2} pressure.
We then pick five random snapshots from the resulting trajectory, remove the hydrogen, fix the Pd--Au sublattice, and run \gls{mc} simulations at \unit[300]{K} with \gls{sgc} flips on the H--vacancy sublattice only, using a wide range of hydrogen chemical potentials.

\begin{figure}
    \centering
    \includegraphics{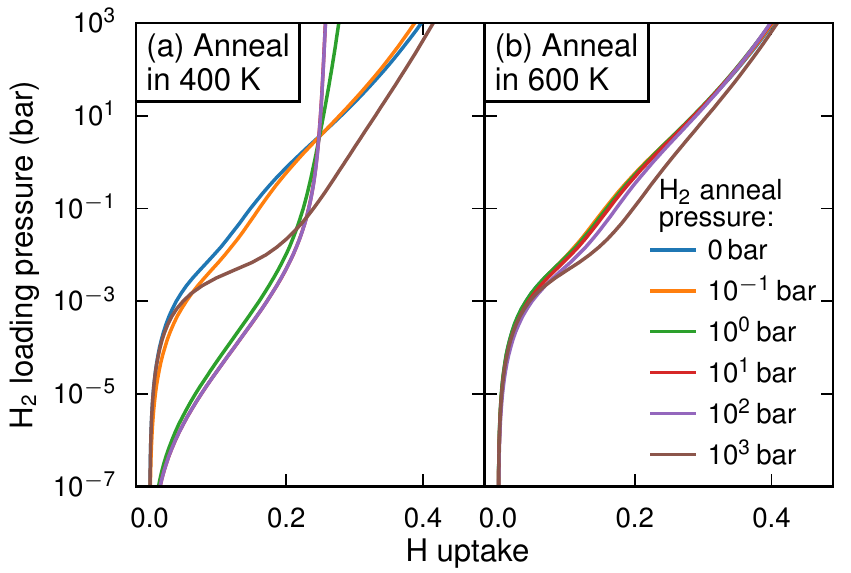}
    \includegraphics{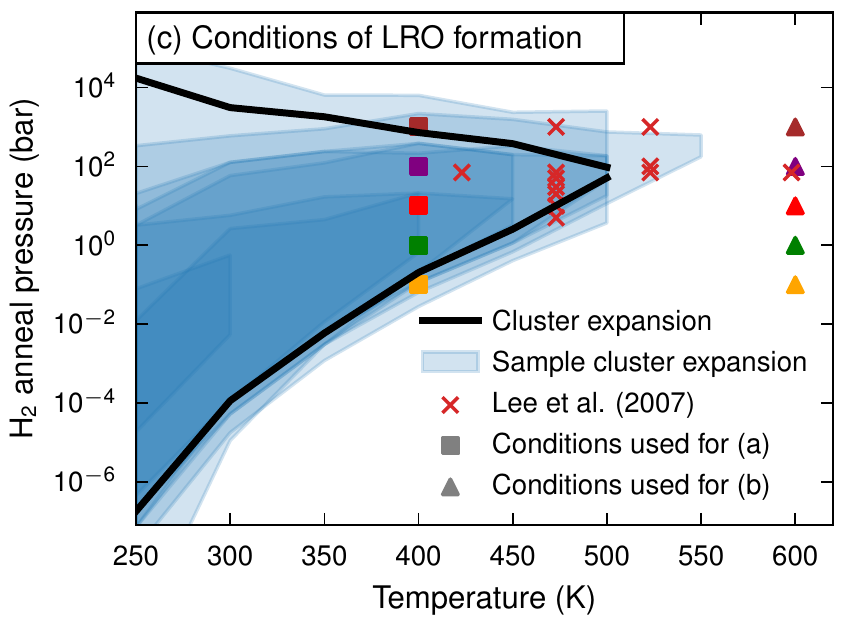}
    \caption{
    Impact of annealing conditions on absorption of H in Pd--Au with 25\% Au at \unit[300]{K}.
    After annealing in \unit[400]{K} (a), the isotherms exhibit a markedly different behavior if the \ce{H2} pressure during annealing was between 1 and \unit[100]{bar} (the red \unit[10]{bar} isotherm is hidden under the green \unit[1]{bar} isotherm).
    After annealing in \unit[600]{K} (b), the isotherms are almost identical regardless of \ce{H2} exposure during annealing.
    The origin of the change in isotherm in (a) is the formation of LRO under certain conditions as quantified in (c).
    The CE predicts that LRO will form in full equilibrium when Pd--Au with 25\% Au is subjected to \ce{H2} pressure and temperature corresponding to the area between the black lines.
    The corresponding area predicted with ten sampled CEs is indicated with transparent blue, one per CE.
    Darker color thus means more \glspl{ce} predict order formation at that point.
    Out of these ten \glspl{ce}, two have a critical temperature too low to be visible in this figure.
    The conditions investigated by Lee \textit{et al.} \cite{LeeNohFla07} with signs of order formation in Pd--Au with 19\% Au are indicated with red crosses.
    Colored squares and triangles indicate the annealing conditions for isotherms in (a) and (b), respectively.
    }
    \label{fig:solubility-changes}
\end{figure}

Inspection of \unit[300]{K} isotherms with Pd--Au annealed in \unit[400]{K} (\autoref{fig:solubility-changes}a) reveals that the hydrogen concentration depends strongly on the conditions during annealing.
When annealed in pressures of $\unit[10^{-1}]{bar}$ or lower (blue and orange lines), the isotherms behave as in the random or para-equilibrium case, with an almost linear isotherm (when plotted on a logarithmic scale).
For higher pressures, however, the isotherms are markedly different.
After annealing in 1, 10 or \unit[100]{bar}, the uptake of hydrogen at low pressures is much higher, after which the concentration of hydrogen stays at about 25\% up to very high pressures, meaning that at sufficiently high pressures, the hydrogen content is in fact higher in Pd--Au samples that were annealed in the absence of hydrogen.
The difference between these two kinds of isotherms is that annealing in 1--\unit[100]{bar} induces L1$_2$ \gls{lro}.
If the \ce{H2} pressure is raised to \unit[1,000]{bar}, the ordered phase no longer forms, and the isotherm (brown line in \autoref{fig:solubility-changes}a) becomes more similar to the low-pressure isotherms, although the hydrogen uptake is significantly higher if the \ce{H2} pressure is above a few millibar.

The ordered L1$_2$ phase is thus clearly distinguishable from the ones lacking \gls{lro} already from the isotherms.
These findings are consistent with Lee \textit{et al.} \cite{LeeNohFla07}.
When annealing in \unit[600]{K}, on the other hand, the temperature is too high for any \gls{lro} to emerge, and the \unit[600]{K} isotherms are essentially identical regardless of annealing pressure (\autoref{fig:solubility-changes}b).
It is expected that much higher pressures are required to impact the system at \unit[600]{K} compared to \unit[400]{K}, because at constant pressure when the temperature goes up, the hydrogen content in the material goes down.
Nevertheless, when the annealing pressure is \unit[100]{bar}, the hydrogen content in the system is about 14\% during annealing, but the \unit[300]{K} isotherm is still virtually unaffected.
Only when the annealing pressure reaches \unit[1,000]{bar} (leading to approximately 26\% hydrogen in the system during annealing), is the \unit[300]{K} isotherm clearly distinguishable from the isotherm of Pd--Au annealed in vacuum, and even then the difference is small.

This picture emerges more comprehensively if we study the hydrogen absorbed at \unit[300]{K} and specific partial pressures (\autoref{sfig:solubility-changes}): the content of hydrogen is virtually independent of annealing conditions as long as the phase transition to the L1$_2$ phase does not occur.
Very high annealing pressures, on the order of 10--\unit[1000]{bar}, are required to achieve a significant impact on the content of absorbed hydrogen unless the ordered L1$_2$ phase is formed.

It is worth stressing that while hydrogen uptake at moderate pressure is enhanced by \gls{lro} formation, we observe a different trend in \autoref{sec:random-para-results}; \gls{sro} formation decreases the hydrogen uptake unless the pressure is very low.
It is thus advisable not to speak in too general terms about the impact of chemical order on hydrogen solubility, because it depends on the details of the chemical order as well as the \ce{H2} pressure at which the solubility is assessed.

Given that the chemical ordering of the L1$_2$ phase thus has a significant impact on the nature of absorption of \ce{H2} in Pd--Au, it may have a notable impact on any utilization thereof.
We therefore estimate the conditions during which this long-range ordered phase will form (\autoref{fig:solubility-changes}c).
Our \gls{ce} indicates that the ordered structure starts to form below approximately \unit[500]{K} with a \ce{H2} pressure between approximately 50 and \unit[100]{bar}.
The \ce{H2} pressure range where \gls{lro} forms then widens quickly as temperature is decreased, but slower kinetics will of course inhibit order formation at too low temperatures.
Lee \textit{et al.} \cite{LeeNohFla07}, who studied Pd--Au with 19\% Au, assumed that distinct isotherms are a fingerprint of \gls{lro}, and while the purpose of their study was not to map out the conditions under which order emerges, their observations are qualitatively consistent with our \gls{ce} (red crosses in \autoref{fig:solubility-changes}c), but with a higher critical temperature (with \gls{lro} persisting up to at least \unit[598]{K}).
It thus seems likely that our \gls{ce} underestimates the critical temperature by at least \unit[100]{K}.
By repeating the same calculations with 10 different \glspl{ce} (grey areas in \autoref{fig:solubility-changes}c), we observe qualitatively the same behavior but find that both the critical temperature and the pressure range are very sensitive to small variations in the \glspl{eci}.
Although none of the sampled \glspl{ce} exhibit a critical temperature above \unit[600]{K}, it should be clear that small errors not captured by the \gls{ce} approach (such as neglect of vibrations or the choice of exchange-correlation functional used to calculate the training data) may cause an error of this magnitude.

\section{Conclusions}
We have comprehensively investigated the impact of chemical order on hydrogenation of Pd--Au alloys using a computational approach based on \gls{dft} calculations, \gls{ce} models, and \gls{mc} simulations.
Although relatively large relaxations from octahedral hydrogen sites hamper the ability of the \gls{ce} models to exactly reproduce formation energies calculated with \gls{dft}, we found that our \gls{ce} reproduced thermodynamic properties of Pd--Au hydrides that are well-established experimentally, with a quantitative agreement on par with what can be expected from a \gls{ce} approach.
This applies especially given the large impact of small changes in the \glspl{eci}, which is likely always inherent in this approach or indeed any other approach to derive phase diagrams from interatomic potentials or first-principles data.
Also, we acknowledge that zero-point energy as well as phonons can sometimes play an important role in hydrides due to the low mass of hydrogen, but inclusion of these effects would have been computationally prohibitive.
The general impact of these effects may warrant future studies and may explain some of the quantitative disagreement with experiment.

Our results provide a rationale for the experimental observation that absorption/desorption isotherms are relatively stable over time; as long as the L1$_2$ phase does not form, isotherms will stay similar even if chemical order changes.
This is manifested by the similarity between isotherms in random and para-equilibrium.
Our results predict that a small reduction in the ability to absorb hydrogen may be observed if Pd--Au annealed at a very high temperature is allowed to reach equilibrium at a much lower temperature.

Under long-term exposure to hydrogen, however, the situation changes substantially.
The emergence of the ordered L1$_2$ phase introduces complexity in the phase diagram, and Pd--Au stored at room temperature and, say, \unit[1]{mbar} \ce{H2} may over time exhibit very different absorption isotherms, as the result of emerging \gls{lro}.
The hydrogen content in air is of course orders of magnitude lower than what we predict is required for this phase to form, but repeated exposure to high \ce{H2} pressures may result in formation of \gls{lro} and an altered isotherm.

Does the \gls{lro} vanish when the hydrogen is removed?
This question is difficult to answer, because from a kinetic point of view, our simulation approach is not able to describe the stability of the ordered phase in absence of hydrogen (or vice versa).
We do predict that the L1$_2$ phase is the ground state at this composition also in absence of hydrogen, but its critical temperature is well below \unit[300]{K}.
It may, however, be noted that there are (hydrogen-free) Pd--Au phase diagrams in the literature where this phase is expected to form above room temperature \cite{MatNagKak66, KawInoOga71, BerLeg20}, that is, even without exposure to hydrogen.
Although the existence of this phase, let alone its critical temperature, remains debated, it seems reasonable to assume that the L1$_2$ phase might be fairly stable at room temperature even after hydrogen is removed.

The emergence of a different isotherm upon ordering might constitute a challenge as the sensor would have a different readout at the same \ce{H2} pressure.
It may, however, also present an opportunity.
The ordered phase absorbs significantly more hydrogen at low pressures.
A sensor consisting of the ordered phase is thus likely (depending on the nature of the readout) to be significantly more sensitive to small changes in \ce{H2} pressure.
If the phase is sufficiently stable, this may provide a relatively simple way to improve a Pd--Au hydrogen sensor---annealing in hydrogen to boost sensitivity.


\begin{acknowledgments}
This work was funded by the Knut and Alice Wallen\-berg Foundation (grant numbers 2014.0226, 2015.0055), the Swedish Research Council (grant number 2018-06482), and the Swedish Foundation for Strategic Research (grant number RMA15-0052).
The computations were enabled by resources provided by the Swedish National Infrastructure for Computing (SNIC) at NSC, C3SE and PDC partially funded by the Swedish Research Council (grant number 2018-05973).
We thank Dr. Jonatan W{\aa}rdh for helpful discussions.
\end{acknowledgments}

\end{document}